\documentclass[ floatfix,twocolumn]{revtex4}
\usepackage{amsmath}
\usepackage{color}
\usepackage{amssymb}
\usepackage{graphicx}
\usepackage[utf8]{inputenc}
\usepackage{textcomp}
\usepackage{color}

\begin{document}

\title{Metamagnetic  phase transition in ferromagnetic superconductor URhGe}

\author{V.P.Mineev$^{1,2\footnote{E-mail: vladimir.mineev@cea.fr}}$}
\affiliation{$^1$Universite Grenoble Alpes, CEA, IRIG, PHELIQS, F-38000 Grenoble, France\\
$^2$Landau Institute for Theoretical Physics, 142432 Chernogolovka, Russia}

\begin{abstract}
Ferromagnetic superconductor URhGe  has orthorhombic structure and possesses  
spontaneous magnetisation along the c-axis.   Magnetic field directed along the $b$-axis suppresses 
ferromagnetism in $c$-direction and leads to a metamagnetic transition  into polarised paramagnetic state in the $b$-direction.
The theory of these phenomena based on the specific magnetic anisotropy of this material  in $(b,c)$ plane is given.
Line of the first order metamagnetic transition ends at a critical point.
The Van der Waals - type description  of  behaviour of physical properties near this point is developed.
The triplet superconducting state destroyed by orbital effect is recreated in vicinity of the transition. 
It is shown that the reentrance of superconductivity  is caused by the sharp increase  of magnetic  susceptibility in $b$ direction near the metamagnetic transition. The specific behaviour of the upper critical field in direction of spontaneous magnetisation in UCoGe and in UGe$_2$ related to the field dependence of magnetic susceptibility is discussed. 
\end{abstract}

\date{\today}
\maketitle
\section{Introduction}

Investigations of uranium superconducting ferromagnets UGe$_2$, URhGe and UCoGe  continue attract attention mostly due to the quite unusual nature of  its superconducting states created by the magnetic fluctuations (see  the recent experimental \cite{Flouquet2019} and theoretical \cite{Mineev2016} reviews and references therein). They have orthorhombic crystal structure and  the anisotropic magnetic properties. The spontaneous magnetisation is directed along $a$ axis in UGe$_2$ and along $c$-axis in URhGe and UCoGe.  The ferromagnetic state in the two last materials is suppressed by the external magnetic field 
$H_y$ directed along  $b$ crystallographic direction. In URhGe at field $H_y=H_{cr}\approx 12$ T  the second order phase transition to ferromagnetic state is transformed to  the transition of the first order \cite{Levy2005}. The superconducting state suppressed \cite{Hardy2005} in much smaller fields $H_y\approx 2$ T is reappeared in vicinity of the first order transition in field interval $ (9,13)$T. The phenomenological theory of this phenomenon has been developed in Ref.5 (see also \cite{Mineev2016}). According to this theory
the state arising in fields above the suppression of spontaneous magnetisation in $c$-direction is the paramagnetic state.

 There was established, however, \cite{Levy2005,Hardy2011,Nakamura2017} that in fields above  $H_{cr}$ 
  the magnetisation along $b$ direction looks like it has field independent "spontaneous" component
\begin{equation}
M_y=M_{y0}+\chi_yH_y.
\end{equation}
This state is called  polarised paramagnetic state. The formation of this state is related with so called 
metamagnetic transition observed in several heavy-fermion compounds (see the paper \cite{YAoki1998}  and the more recent publication \cite{DAoki2011} and references therein).
To take into account the  formation  of polarised paramagnetic state
 one must introduce definite modifications in the treatment performed in  \cite{Mineev2015}.
 Here I present the corresponding derivation.

 The paper is organized as follows. In the Section II after the brief reminder  of results of the paper \cite{Mineev2015} the description of the metamagnetic transition is presented. It is based on the specific phenomenon  of  magnetic anisotropy in URhGe obtained with a local spin-density approximation calculations
 by Alexander Shick \cite{Shick2002}. 
 After the general consideration of the metamagnetic transition the modifications introduced by the uniaxial stress are considered. 
 Then  the Van der Waals - type theory of phenomena  near the metamagnetic critical point is developed and some physical properties are discussed.
 
 The phenomenon of the reentrant superconducting state is explained in the  Section III. It is shown that the recreation of superconductivity is caused
by  the sharp increase in the magnetic susceptibility 
 \cite{Nakamura2017}
 in $b$  direction near the metamagnetic transition. This Section also contains the qualitative description of the  specific behaviour of the upper critical field in direction of spontaneous magnetisation in UCoGe and in UGe$_2$ related to the field dependence of magnetic susceptibility.

 The Conclusion contains the summary of the results.

\section{Metamagnetic transition in ${\bf URhGe}$}

As in the previous publications (\cite{Mineev2016,Mineev2015})  I shall use $x, y, z$ as the coordinates
 pinned to the corresponding
crystallographic directions
$a, b, c$.
The Landau free energy of an 
 orthorhombic ferromagnet 
in
magnetic field ${\bf H}({\bf r})=H_y\hat y$ is
\begin{eqnarray}
F=\alpha_{z}M_{z}^{2}+\beta_{z}M_{z}^{4}+\delta_zM_z^6~~~~~~~~~~~~~~~~~\nonumber\\
 +\alpha_{y}M_{y}^{2} +\beta_{y}M_{y}^{4}+\delta_yM_y^6
 +\beta_{yz}M_{z}^{2}M_{y}^{2}
 -H_yM_y,
\label{F11}
\end{eqnarray}
Here 
\begin{equation}
\alpha_{z}=\alpha_{z0}(T-T^c_{c0}),~~~ \alpha_y>0,
\end{equation}
and
I bear in mind  the terms of the sixth order in powers of $M_z$,  $M_y $ 
and also the fact that in the absence of a field in $x$-direction the magnetisation
along hard $x$-direction $M_x=0$. 

\subsection {Transition ferro-para}

Let us remind first the treatment developed in Ref.5 undertaken in the assumption $\beta_y>0$. Then 
in constant magnetic field ${\bf H}=H_y\hat y$
the equilibrium magnetisation projection along the $y$ direction
\begin{equation}
M_y\approx\frac{H_y}{2(\alpha_y+\beta_{yz}M_z^2)}
\label{My}
\end{equation} 
is obtained by minimisation of free energy (\ref{F11}) in respect of $M_y$ neglecting the higher order terms.
Substituting this expression back to (\ref{F11})
we obtain
\begin{eqnarray}
F=\alpha_{z}M_{z}^{2}
+\beta_{z}M_{z}^{4}+\delta_zM_z^6-\frac{1}{4}\frac{H_y^2}{\alpha_y+\beta_{yz}M_z^2},
\label{F1}
\end{eqnarray}
that gives after expansion of the denominator in the last term, 
\begin{equation}
F=-\frac{H_y^2}{4\alpha_y}+\tilde\alpha_{z}M_{z}^{2}
+\tilde\beta_{z}M_{z}^{4}+\tilde\delta_zM_z^6+\dots,
\label{F2}
\end{equation}
where
\begin{eqnarray}
&\tilde\alpha_{z}=\alpha_{z0}(T-T_{c0})+\frac{\beta_{yz}H_y^2}{4\alpha_y^2},\\
&\tilde\beta_{z}=\beta_z-\frac{\beta_{yz}}{\alpha_y}\frac{\beta_{yz}H_y^2}{4\alpha_y^2}\label{beta},\\
&\tilde\delta_{z}=\delta_z+\frac{\beta_{yz}^2}{\alpha_y^2}\frac{\beta_{yz}H_y^2}{4\alpha_y^2}.
\end{eqnarray}
Thus, in a magnetic field perpendicular to the direction of spontaneous magnetization  the Curie temperature decreases as
\begin{equation}
T_c=T_c(H_y)=T_{c0}-\frac{\beta_{yz}H_y^2}{4\alpha_y^2\alpha_{z0}}.
\label{Cur}
\end{equation}
The coefficient $\tilde\beta_z$ also decreases with $H_y$ and reaches  zero at
\begin{equation}
H_y=H^\star=\frac{2\alpha_y^{3/2}\beta_z^{1/2}}{\beta_{yz}}.
\label{Htcr}
\end{equation}
At this field under fulfilment the  condition,
\begin{equation}
\frac{\alpha_{z0}\beta_{yz}T_{c0}}{\alpha_y\beta_z}>1
\end{equation}
the Curie temperature (\ref{Cur}) is still positive and 
the phase transition from  the ferromagnetic  to the paramagnetic state becomes the transition of the first order
(Fig 1a). The point $(H^\star,T_c(H^\star))$ on the line  paramagnet-ferromagnet phase transition is a tricritical point.
The qualitative field dependences of the normalised Curie temperature $t_c(H_y)=\frac{T_c(H_y)}{T_{c0}}$ and  $b(H_y)=\frac{\tilde\beta_{z}}{\beta_z}$ are plotted in Fig 1a. 

On the line of the first order phase transition from the ferromagnet to the paramagnet state the $M_z$ component of magnetisation drops  
from  $M^{\star}_z$ to zero  \cite{Mineev2016}. The $M_y$ component jumps from $M_y\approx\frac{H^\star}{2(\alpha_y+\beta_{yz}M^{\star 2}_z)}$ to $M_y\approx\frac{H^\star}{2\alpha_y}$. Then at fields $H_y>H^\star$ 
\begin{equation}
M_y\approx\frac{H_y}{2\alpha_y}
\end{equation}
proportional to the external field. This contradicts
experimental observations \cite{Levy2005, Hardy2011, Nakamura2017} which  demonstrate the presence of a "spontaneous" part of magnetization in the field above the transition in accordance with Eq.(1).

\subsection{Transition ferro-polarised para}

The part of free energy depending on $M_y$ 
\begin{equation}
F_y=
\alpha_{y}M_{y}^{2} +\beta_{y}M_{y}^{4}+\delta_yM_y^6
 +\beta_{yz}M_{z}^{2}M_{y}^{2}
 -H_yM_y,
\end{equation}
can be used also  far from the transition to the ferromagnetic state in the temperature region where $M_z$ is not small.
The important fact  obtained with the local spin-density approximation calculations \cite{Shick2002} is that
the coefficient $\beta_y<0$. 
In frame of isotropic Fermi liquid model the negativeness  of the fourth order term in the expansion of the free energy in power of magnetic moment is usually ascribed to the peculiar behaviour of the electron density of states (see the review \cite{Levitin1988} and referencies therein).
In the orthorhombic URhGe  this  specific magnetocrystalline anisotropy reveals itself   in the system of magnetic moments  localised on the uranium atoms \cite{Sanchez2017}.

The  $M_y$ component of magnetisation  is determined  by the equation
\begin{equation}
2\tilde\alpha_{y}M_{y} +4\beta_{y}M_{y}^{3}+6\delta_yM_y^5
 =H_y,
\end{equation}
where
\begin{equation}
\tilde\alpha_y=\alpha_y+\beta_{yz}M_z^2.
\end{equation}
Taking into account the third order term we obtain 
\begin{equation}
M_y\approx \frac{H_y}{2\tilde\alpha_y}-\frac{\beta_yH_y^3}{2\tilde\alpha_y^4}.
\end{equation}
The coefficient $\beta_y<0$ and we see that the increase of magnetisation occurs faster than it was according to Eq.(\ref{My}). 

The  shape of $M_y(H_y)$  depends from the temperature and pressure dependence of coefficients $\alpha_y,\beta_y,\delta_y$. 
In particular, the coefficient  $\tilde\alpha_y(T)$ is decreasing function of temperature   and  at temperature decrease  
the field dependence of $M_y$  transfers
from the monotonous growth taking place at 
$
\beta_y^2<\frac{5}{3} \tilde\alpha_y\delta_y
$
 to the S-shape dependence realising at 
$
\beta_y^2>\frac{5}{3} \tilde\alpha_y\delta_y$.  This transformation occurs at some temperature $T_{cr}$ such that in 
the dependence $H_y(M_y)$ appears an
inflection point. It is
 determined by the equations 
 \begin{equation}
 \frac{\partial H_y}{\partial M_y}=0,~~~~~\frac{\partial^2 H_y}{\partial M_y^2}=0
 \end{equation}
  having common solution
\begin{equation}
M_{cr}^2=-\frac{\beta_y}{5\delta_y},
\label{55}
\end{equation}
at $\beta_y^2=\frac{5}{3} \tilde\alpha_y\delta_y$.
The corresponding critical field is
 \begin{equation}
 H_{cr}=H_y(M_{cr})=\frac{16}{5\sqrt{3}}\frac{\tilde\alpha_y^{3/2}}{|\beta_y|^{1/2}}.
 \label{66}
 \end{equation}

At $T<T_{cr}$ the inequality
\begin{equation}
\beta_y^2>\frac{5}{3} \tilde\alpha_y\delta_y
\label{iq}
\end{equation}
is realised and
the equation $\frac{\partial H_y}{\partial M_y}=0$ acquires two real solutions, hence,
 the field dependence of $M_y$ acquires the S-shape plotted at Fig.1b. 
 Equilibrium transition from the lower  to the upper part of the curve $M_y(H_y)$ corresponds to a vertical line connecting the points $M_1$ and $M_2$ defined by the Maxwell rule $\int_1^2M(H)dH=0$. The integration is performed along the curve $M_y(H_y)$.
The $M_y$ component of magnetisation jumps from $M_1$ to $M_2$ (see Fig1b).

At temperatures above $T_{cr}$ the jump transforms into the crossover  which is the temperature-field region characterised by the fast growth $M_y$.
The lower boundary of this region roughly coincides with the Curie temperature (see Fig2.).
The Curie temperature decreasing with growth of magnetisation $M_y$
 \begin{equation}
 T_c(H_{y})=T_{c0}-\frac{\beta_{yz}M_y^2}{\alpha_{z0}}
 \label{eq}
 \end{equation}
 falls down to zero or even to negative value at sharp increase of  $M_y$
  in vicinity of the critical field $H_{cr}$
 and the ferromagnetic order along $z$-direction disappears. 
 Thus, at $T<T_{cr}$ and $H_y=H_{cr}$ we have the phase transition of the first order from the ferromagnetic state with spontaneous magnetisation along $z$-direction to the polarised paramagnetic  state with induced magnetisation along $y$-direction (Fig2).

The described  jump-like transition is realised in  the cylindrical specimen in the magnetic field parallel to the cylinder axis. In specimens of the arbitrary shape with demagnetisation factor $n$  the transition occurs in some field interval where the specimen is filled by the domains with different magnetisation.

When the critical field $H_{cr}$ is smaller than the critical field of  transition ferro-para $H^\star$, 
the ferro-para transition, discussed in the previous section does not occurs. 

 At  $T<T_{cr}$   in fields  $H_y$  exceeding $H_{cr}$  the field dependence of $M_y$ component of magnetisation 
 behaves in accordance with  Eq.(1) corresponding to the experimental observations. 
 
 \subsection{Uniaxial stress effects}
 
 It is known that a hydrostatic pressure applied to URhGe crystals stimulate ferromagnetism and at the same time suppresses  the superconducting state \cite{F.Hardy2005} and the reentrant superconducting state \cite{Miyake2009} as well. The later is also shifted to a bit higher field interval. On the contrary, the uniaxial stress  along $b$-direction suppresses the ferromagnetism decreasing the Curie temperature and stimulates the superconducting state so strongly that it leads to the coalescence of the superconducting and reentrant superconducting regions in the $(H_y,T)$ phase diagram \cite{Braithwaite2016}. The phenomelogical description of these phenomena was undertaken in the paper\cite{Mineev2017}. There was shown that   both coefficients $\alpha_z$ and $\alpha_y$ in the Landau free energy Eq.(\ref{F11}) acquire the linear uniaxial pressure dependence
 \begin{eqnarray}
\alpha_{z}(P_y)=\alpha_{z0}(T-T_{c0})+A_zP_y,\\ 
\alpha_y(P_y)=\alpha_{y}-|A_y|P_y~~~~~
\end{eqnarray}
corresponding to the moderate uniaxial pressure suppression of the Curie temperature 
 \begin{equation}
 T_{c}(P_y)= T_{c0}-\frac{A_zP_y}{\alpha_{z0}},
 \end{equation}
 reported in \cite{Braithwaite2016} in the absence of an external field.
 However, under the external field along $y$-direction the drop of the Curie temperature  Eq.(\ref{Cur}) is  accelerated
 \begin{eqnarray}
T_c(H_y,P_y))\approx T_{c0}-\frac{A_zP_y}{\alpha_{z0}}
-\frac{\beta_{yz}H_y^2}{4(\alpha_y(P_y))^2\alpha_{z0}}
\end{eqnarray}
in correspondence with the observed behaviour.
Moreover, the uniaxial stress causes  strong decrease of the critical field  Eq.(\ref{66}) 
 \begin{equation}
 H_{cr}=H_y(M_{cr})=\frac{16}{5\sqrt{3}}
 \frac{(\tilde\alpha_y(P_y))^{3/2}}{|\beta_y|^{1/2}}.
 \end{equation}

 \subsection{Van der Waals-type theory near the critical point}

The critical end point temperature for the first order transition in URhGe is $T_{cr}=4$ K and the critical hield is $H_{cr}=12 T$.
 Let us  expand the function $H_y(M_y)$ at temperature slightly deviating from critical temperature $T=T_{cr}+t$ and the magnetisation near its critical value  $M_y=M_{cr}+m$. We have 
 \begin{eqnarray}
h= H_y-H_{cr}=bt+\left [ \frac{\partial H_y}{\partial M_y}|_{t=0} +2at\right ]m\nonumber\\+ \frac{1}{2} \frac{\partial^2 H_y}{\partial M_y^2}|_{t=0}m^2+
  \frac{1}{6}\frac{\partial^3 H_y}{\partial M_y^3}|_{t=0}m^3,
 \end{eqnarray}
 Here, we neglected   by the temperature dependence of the  second and the third order terms. Taking into account that $ \frac{\partial H_y}{\partial M_y}|_{t=0}=\frac{\partial^2 H_y}{\partial M_y^2}|_{t=0}=0$
 we obtain  
   \begin{equation}
 h=bt+2atm+ 4Bm^3,
 \label{vdW}
 \end{equation}
 which obviously corresponds to the expansion of pressure $p=P-P_{cr}$ in powers of density $\eta=n-n_{cr}$ 
near the Van der Waals critical point  \cite{StatPhys}.

At $t<0$ according to the Maxwell rule the magnetisation densities of two phases in equilibrium with each other are:
\begin{equation}
m_{2}=-m_{1}=\sqrt{\frac{-at}{2B}}.
\end{equation}
 
 The line of phase equilibrium between the two phases below and above the transition is given by the equation 
 \begin{equation}
 h=bt, ~t<0. 
 \end{equation}

\subsubsection{Specific heat}

The specific heat at fixed external field(see \cite{StatPhys}) is
\begin{equation}
C_{h}\propto T\frac{\left (\frac{\partial h}{\partial t}\right )^2_m}{\left (\frac{\partial h}{\partial m}\right )_t}.
\end{equation}
Then,  using Eq.(\ref{vdW})   
  we obtain 
\begin{equation}
C_{h}\propto\frac {b^2T}{2at+12Bm^2}.
\end{equation}
Thus,  the contribution to heat capacity according to the equation of state  (\ref{vdW})   near the critical point  grows so long $m^2$ decreases till to $m^2_{1}$ and then begins to fall when $m^2$ increases starting from $m^2_{2}$ (see Fig3). This is the contribution
to the specific heat 
of the whole system 
and cannot be directly attributed to the specific heat of itinerant electrons proportional to  the electron effective mass. 

The low temperature behaviour of the URhGe specific heat in magnetic field has not been established by a direct measurement but was derived \cite{Hardy2011} by the application
 of the Maxwell relation $\left (\frac{\partial S}{\partial H_y}\right )_T=\left (\frac{\partial M_y}{\partial T}\right )_{H_y}$ from the temperature dependence 
of the magnetisation $M_y(T,H_y)$  in the fixed field.  The changes of the ratio $C(T)/T$ have been ascribed to the the electron effective mass dependence from magnetic field \cite{Miyake2008,Hardy2011}. This was done in the assumption that URhGe is a weak itinerant ferromagnet, in other words, all the low temperature degrees of freedom in this material belong to the itinerant electron subsystem. As we already mentioned above,  the strong magnetic anisotropy of this material \cite{Shick2002} points on the importance of the magnetic degrees of freedom localised on the uranium ions  and related with crystal field levels \cite{Sanchez2017,Mineev2016}.

\subsubsection{Resistivity}

The magnetic field dependence of effective mass  was also found \cite{Miyake2008,Gourgout2016} by the aplication
 the Kadowaki-Woods relation $A(H_y)\propto (m^\star)^2$
where coefficient $A$ is a pre-factor in the low-temperature dependence of resistivity $\rho=\rho_0+AT^2$. 

 The $A(H_y)$ behaviour  
 is determined by the processes of inelastic electron-electron scattering which in the multi-band metals interfere with scattering on impurities (see fi \cite{Keyes1958,Gantmakher, Appel1978,Murzin,Pal2012}) and on magnetic excitations with field dependent spectrum. The non-spherical shape of the Fermi surface sheets  and the screening of el-el Coulomb interaction can  introduce deviations from $T^2$ resistivity dependence. So, the physical meaning of the coefficient $A(H_y)$ behaviour 
 is not so transparent and its relationship with the electron effective mass is questionable. 
 
 One can also note, that
the temperature fit of the experimental data was done in very narrow temperature interval and  the  $T^2$ temperature dependence claimed in  \cite{Gourgout2016}  seems somewhat unreliable. Compare  with  the results reported in \cite{Prokes2002,F.Hardy2005}. 
  
\subsubsection{Correlation function}

The correlation function of fluctuations of the magnetisation density $m$  near the critical point at $t<0$ behaves similar to the specific heat  \cite{StatPhys}
\begin{equation}
\varphi({\bf k})= \frac {T}{2(at+6Bm^2+ \gamma_{ij}k_ik_j)}.
\end{equation}
This is in correspondence  with a marked increase of the NMR relaxation rate $1/T_2$ with field $H_y$ increasing toward 12 T reported in \cite{Kotegawa2015, Tokunaga2015}.

\section{ Phase transition to superconducting state}

The superconducting state in URhGe is completely suppressed by the magnetic field  $H_{c2}(T=0)\approx 2$ T in $y$-direction due to the orbital depairing effect.
Then superconductivity recovers  in the field interval $9-13$ T around the critical field $H_{cr}\approx 12$ T of the transition of the first order 
 from the ferromagnetic state with spontaneous magnetization along z-direction to the  state with induced magnetization along $y$-direction. Evidently such type behaviour is possible if the magnetic field somehow stimulates the pairing interaction surmounting the orbital depairing effect.

  In numerous publications starting from the paper by A.Miyake et al \cite{Miyake2008}  the treatment of this phenomenon was related with the assumption of an enhancement of electron effective mass $m^\star=m(1+\lambda)$   leading to the enhancement of pairing interaction  and consequently of the temperature of transition to superconducting state
 according to 
the Mc-Millan-like formula \cite{McMillan1968}
\begin{equation}
T_{sc}\approx \epsilon\exp\left (-\frac{1+\lambda}{\lambda}\right )
\end{equation}
derived  in the paper \cite{Fay1980} for the superconducting state with $p$-pairing  in an  itinerant isotropic ferromagnetic metal. 
Similar to the liquid He-3 in this model there are two independent phase transition to the superconducting state in the subsystems with spin-up and spin-down electrons.  
The constant $\lambda$  determined by
 the Hubbard four-fermion interaction \cite{Fay1980,BE} increases 
 as we approach but not too much close to ferromagnetic  instability. In frame of this model the question of why the growth of the magnetic field $H_y$ 
approaches the ferromagnetic transition remains unanswered.

The following development of this type approach has been undertaken by Yu.Sherkunov and co-authors \cite{Chubukov2018}. The reentrant superconductivity and mass enhancement  have been associated with the Lifshitz transition \cite{Yelland2011} which occurs in one of the bands in a finite magnetic field stimulating the splitting of spin-up and spin-down bands. There was established modest enhancement of the transition critical temperature in the field about 10 T. 
Thus, the model can claim to the  qualitative explanation of the  superconducting state reentrance.
However, it should be noted that  the measured \cite{Yelland2011}
 quasiparticle mass in the corresponding band does not increase but decreases and remains finite, implying that the Fermi velocity vanishes due to the collapse of the Fermi wave vector. The cross-section of the Fermi surface  of this band corresponds to 7\%   of the Brillouin zone area.
 Thus, the reentrance of superconductivity is hardly could be associated with the observed Lifshitz transition.

The models \cite{Fay1980, Chubukov2018}  describe the physics of pure itinerant electron subsystem. Such a treatment is approved in application  to the 
$^3$He Fermi-liquid. The measurements by x-ray magnetic circular dichroism \cite{Sanchez2017} point to the local nature  of the URhGe ferromagnetism.  Namely, the comparison of the total uranium moment $\mu_{tot}^U$ to the total magnetisation $M_{tot}$ at different magnitude and  direction of magnetic field indicates that the uranium ions dominate the magnetism of URhGe. The same is true also in the parent compound UCoGe \cite{Rogalev2015}.
So, the magnetic susceptibility $\chi_{ij}({\bf q}, \omega)$ is mostly determined by the localised moments subsystem.
Hence,
an approach based on the exchange interaction between conduction electrons and magnetic moments localised on uranium atoms seems more appropriate. This type theory
has been developed in the paper  by Hattori and Tsunetsugu \cite{Hattori2013}. Here, there will be undertaken  an other approach  allowing explicitly  take into account  the enhancement of magnetic susceptibility
 near the metamagnetic
 transition from the ferromagnet state with spontaneous magnetisation along the $c$-axis to the magnetic state polarised along the $b$-axis.

Using the standard functional-integral representation of the
partition function of the system (see fi \cite{Karchev2003}), we obtain the following term in
the fermionic action describing an effective two-particle
interaction between electrons:
\begin{eqnarray}
\label{S}
  S_{int}= -\frac{1}{2}I^2\int dx\,dx'
    S_i(x){\cal D}_{ij}(x-x')S_j(x'),
\end{eqnarray}
where 
$
{\bf S}({\bf r})=\psi^\dagger_\alpha({\bf r})\mbox{\boldmath$\sigma$}_{\alpha\beta}\psi_\beta({\bf r})
$ 
is the operator of the electron spin density,
$x=({\bf r},\tau)$ is a shorthand notation for the coordinates
in real space and the Matsubara time, $\int dx(...)=\int
d{\b r}\int_0^\beta d\tau(...)$, $I$ is the exchange constant of interaction of itinerant electrons with localised magnetic moments, ${\cal D}_{ij}(x-x')$ is the
spin-fluctuation propagator expressed in terms of the dynamical spin susceptibility $\chi_{ij}({\bf q}, \omega)$.

Making use the interaction (\ref{S}) one can calculate the  electron self energy and find the dependence of the electron effective mass from magnetic field as well the temperature of transition to the superconducting state with triplet pairing. 
The energy of electronic excitations in temperature region where the superconducting state is realised
is much smaller than typical energy of magnetic excitations.
Hence, in calculation of the superconducting properties one can neglect the frequency dependence of susceptibility.

\subsection{Upper critical field parallel to $c$-axis in UCoGe}

In application to UCoGe  in magnetic field parallel to direction of spontaneous magnetisation this program has been accomplished in the paper \cite{Mineev2020}.  There
has been considered transition into the  equal-spin pairing  superconducting state in  two-band (spin-up, spin-down) orthorhombic ferromagnetic metal.
According to this paper
in the simplified case of a single-band (say spin-up) equal-spin pairing superconducting state the critical temperature without including the orbital effect of the field is
\begin{equation}
T_{sc}=\epsilon~\exp\left (-\frac {1+\lambda}{\langle N_0({\bf k})\chi_{zz}^u\rangle I^2} \right ),
\label{ct}
\end{equation}
 where, as in the McMillan formula, $1+\lambda$ corresponds to the effective mass renormalisation, whereas the pairing amplitude expressed through the odd in momentum part of static susceptibility 
 $$
  \chi_{zz}^u=\frac{1}{2}\left[\chi_{zz}({\bf k}-{\bf k}^\prime)-\chi_{zz}({\bf k}+{\bf k}^\prime)\right ],
  $$
 which is the main source of the critical temperature dependence from magnetic field.
 Here,
 \begin{equation}
 \chi_{zz}({\bf k})=\frac{1}{\chi_z^{-1}+2\gamma_{ij}k_ik_j},
 \end{equation}
  and $\chi_z=\chi_z(H_z)$ is the $z$-component of susceptibility  in the finite  field $H_z$. Its magnitude at $H_z\to 0$ we will denote $\chi_{z0}$. The angular brackets denote averaging over the Fermi surface and $N_0({\bf k})$ is the angular dependent density of electronic states on the Fermi surface,
 \begin{equation}
 \langle N_0({\bf k})\chi_{zz}^u(H_z)\rangle\approx \frac{2\langle N_0({\bf k})\hat k_z^2\rangle k_F^2 \chi_z}{(2\chi_z)^{-1}+4\gamma_{zz}k_F^2}.
\end{equation} 
The denominator in the exponent of Eq.(\ref{ct})  can be expressed through its value at $H_z\to 0$
\begin{equation}
\frac{\langle N_0({\bf k})\chi_{zz}^u(H_z)\rangle}{ \langle N_0({\bf k})\chi_{zz}^u(H_z\to 0)\rangle}
= \frac{\chi_z}{\chi_{z0}}\frac{1+4(\xi_mk_F)^2}{\frac{\chi_{z0}}{\chi_z}+4(\xi_mk_F)^2}.
\label{es}
 \end{equation}
Here the  product $  2\gamma_{zz}k_F^2\chi_{z0}=(\xi_mk_F)^2$ is expressed through the magnetic coherence length $\xi_m$ which 
near the zero temperature is  of the order several interatomic distances. 

In assumption $(\xi_mk_F)^2\gg 1$ one can rewrite the Eq.(\ref{es}) as
\begin{equation}
\langle N_0({\bf k})\chi_{zz}^u(H_z)\rangle
\approx \frac{\chi_z(H_z)}{\chi_{z0}}
 \langle N_0({\bf k})\chi_{zz}^u(H_z \to 0)\rangle.
\label{et}
 \end{equation}
This very rough estimation  presents the qualitative dependence of exponent in equation (\ref{ct}) from magnetic field.
The longitudinal susceptibility drops with 
the augmentation of magnetic field parallel to the spontaneous magnetisation (see Fig 3 in the paper \cite{Huy2008})
  leading to the suppression of the temperature of transition to the superconducting state  without including the orbital effect according Eq.(\ref{ct}).

  Taking into account the orbital effect one can write the field dependence of critical temperature of transition to the superconducting state in
  the Ginzburg-Landau region
 \begin{equation}
 T_{sc}^{orb}(H)=T_{sc}(H)-\frac{H}{AT_{sc}(H)},
 \end{equation} 
 where $A$ is a constant.
 Thus, the decreasing of $T_{sc}(H)$ with magnetic field causes not only faster drop   but also the peculiar  upward curvature in 
  the critical temperature  $T_{sc}^{orb}(H)$ dependence from magnetic field in correspondence with the experimental data reported in \cite{Beilun2017}.


\subsection{Reentrant superconductivity in URhGe}

In field perpendicular to the spontaneous magnetisation 
the similar approach applied to the simplified single band model  in weak coupling  approximation yields (see Eq.(169) in the review \cite{Mineev2016}) the critical temperature 
 \begin{eqnarray}
T_{sc}\approx~~~~~~~~~~~~~~~~~~~
\nonumber\\ \epsilon~\exp\left (-\frac {1}{[
\langle N_0({\bf k})\chi_{zz}^u\rangle\cos^2\varphi+ 
\langle N_0({\bf k})\chi_{yy}^u\rangle\sin^2\varphi]I^2} \right ),~
\label{nnn}
\end{eqnarray}
where $\tan\varphi=H_y/h$ and $h$ is the exchange field acting on the electron spins. This is the critical temperature of transition to the superconducting state without including the orbital effect.

The orbital effect suppresses the superconducting state and near  the upper critical  field at zero temperature
\begin{equation}
H_{c2y}(T=0)=H_0=cT_{sc}^2
\end{equation}
 the actual critical temperature is
\begin{equation}
T_{sc}^{orb}=a\sqrt{H_0-H_y},
\label{bcs}
\end{equation}
where  $a\sqrt{c}$ is the numerical constant of the order of unity. This is the usual square root BCS dependence of the critical temperature from magnetic field in {\bf low temperature - high field} region such that $T_{sc}^{orb}(H_y=H_0)=0$.  However, in the present case the magnitude $H_0$ itself is a function of the external field $H_y$. Let us look on its behaviour.

Similar to Eq.(\ref{et}) we get
\begin{eqnarray}
\langle N_0({\bf k})\chi_{zz}^u(H_y)\rangle\cos^2\varphi+ 
\langle N_0({\bf k})\chi_{yy}^u(H_y)\rangle\sin^2\varphi\approx\nonumber\\
 \frac{\chi_z(H_y)}{\chi_{z0}}\langle N_0({\bf k})\chi_{zz}^u(H_y\to 0)\rangle\cos^2\varphi~~~~~~~~~~~~~~~~~\nonumber\\+ 
 \frac{\chi_y(H_y)}{\chi_{y0}}\langle N_0({\bf k})\chi_{yy}^u(H_y\to 0)\rangle\sin^2\varphi.~~~~~~~~~~~~
 \label{em}
\end{eqnarray}
Here, $\chi_z(H_y)$ and $\chi_y(H_y)$ are the $z$ and $y$ components of susceptibility in finite field $H_y$ and $\chi_{z0}$ and $\chi_{y0}$ are the corresponding susceptibilities at $H_y\to 0$. Unlike to the Eq.(\ref{et}) the field dependence of the  Eq.(\ref{em})
is not so visible. One can note, however, the different field dependence of two summands in the Eq.(\ref{em}).
\\
(i)  The susceptibility along $z$ direction $\chi_z(H_y)$ increases with magnetic field $H_y$ following to the decreasing of the Curie temperature according to Eq.(\ref{eq}). The growth of susceptibility along z direction at the  approaching  the  field $H_y$ to $H_{cr}$ is confirmed by the field dependence of the NMR  scattering rate $1/T_1$ reported in \cite{Tokunaga2015,Kotegawa2015}.  At the same time, the increase of  $\chi_z(H_y)$ is limited by the decrease of $\cos^2\varphi$. We do not know how fast it is because the magnitude of the exchange field is not known.\\
(ii) As the field approaches to $H_{cr}$ the low temperature susceptibility $\chi_y(H_y)$ has a high delta-function-like peak \cite{Nakamura2017} with magnitude
more than 10 times greater than it is at $H_y\to 0$.
The factor $\sin^2\varphi$ is also increased. This indicates that in URhGe the more important is the second term connected with the metamagnetic transition.

Thus, in vicinity of metamagnetic transition one can expect the increase of the critical temperature 
estimated without including the orbital effect according to Eq.(\ref{nnn}). The radicand in the equation (\ref{bcs})  after being  negative in some field interval acquires the positive value as the field approaches  to $H_{cr}$. The critical temperature Eq.(\ref{bcs})   reaches maximum in vicinity of metamagnetic transition, see Fig2.

Similar arguments  in favour of stimulation superconductivity near the metamagnetic transition in field parallel to $b$ axis can be applied to the 
 discovered recently  other superconducting compound UTe$_2$  \cite{Ran-NatPhys, Knebel2019,Pourret2020} isostructural with URhGe. However, in view of many particular properties of this material we leave this subject for future studies .

In the parent compound UCoGe the metamagnetic transition is absent (at least at $H_y< 40$T) \cite{Knafo2012}. Hence, in this material the unusual temperature dependence  of the upper critical field parallel to $b$ axis  is  probably mostly determined by the first term in the Eq.(\ref{em}).

Near $H_y=H_{cr}$ at temperatures $T<T_{cr}$ the NMR spectrum is composed of two components indicating that the transition is of the first order accompanied by the phase separation \cite{Kotegawa2015}. Thus, in almost whole interval near $H_{cr}$  the superconductivity is developed in mixture of ferromagnetic state with polarisation along $z$ direction and the field polarised state with polarisation along $y$-direction.

\subsection{ Upper critical field near metamagnetic transition in UGe$_2$}

Peculiar example of superconductivity stimulation  in vicinity of metamagnetic transition is realised in the other ferromagnetic compound UGe$_2$.
This material has orthorhombic structure with spontaneous magnetisation directed along $a$ crystallographic direction. The magnetism in UGe$_2$ 
has even more localised nature \cite{Mineev2016,Troc2012,Mineev2013} than in related compounds URhGe and UCoGe.
The superconductivity
exists inside of ferromagnetic  state in the pressure interval  shown in Fig4. Inside of this interval at $P=P_x$ there is a metamagnetic transition  from ferromagnetic state FM1 to ferromagnetic state FM2 characterised by the jump of spontaneous magnetisation from smaller to larger value \cite{Huxley2002}.  At 
a bit higher pressure $P=P_x+\delta P$ the transition from FM1 to FM 2 occurs in a finite magnetic field applied along the direction of spontaneous magnetisation. Near this transition in a finite field  the magnetic susceptibility along the $a$-axis $\chi_a$ strongly increases. Hence, the critical temperature without including the orbital effect
\begin{equation}
T_{sc}=\epsilon~\exp\left (-\frac {1+\lambda}{\langle N_0({\bf k})\chi_{a}^u\rangle I^2} \right ),
\label{ctt}
\end{equation}
 growths up. As result the upper critical field in
$a$ crystallographic direction measured at  $P=P_x+\delta P$ acquires non-monotonic temperature dependence shown in  Fig 5 \cite{Huxley2001,Sheikin2001}. 

It worth to be noted that at pressures far from metamagnetic transition the upper critical field parallel to $a$-direction does not reveal an upward curvature \cite{Huxley2001,Sheikin2001}. This important distinction from the upper critical field behaviour in UCoGe considered in Section IIIA
is related to the difference of susceptibility dependence from magnetic field along spontaneous magnetisation in these two materials. Whereas in UCoGe the susceptibility  $\chi_c$ along $c$-axis is strongly field dependent \cite{Huy2008}, in UGe$_2$ the susceptibility $\chi_a$ along $a$-axis is practically field independent \cite{Huxley2002,Tateiwa2018}.

 \section{Conclusion}
 
We have demonstrated that in the orthorhombic ferromagnet URhGe the ferromagnetic ordering  along $c$-axis suppressed in process of increase of magnetization in the perpendicular $b$-direction induced by the external magnetic field. 
This process is accelerated by the tendency to the metamagnetic transition which occurs at $H_y=H_{cr}=12$ T. The transition of the first order is accompanied by the suppression of the ferromagnetic state with polarisation along $c$-axis and the arising of magnetic state polarised along $b$-xis. The line of first order phase transition is finished at the critical end point with temperature $T=T_{cr}=4$ K. 

 The uniaxial stress along $b$-axis  causing moderate suppression of the Curie temperature in the absence of magnetic field accelerates the Curie temperature drop in finite magnetic field $H_y$ and quite effectively decreases the critical field 
of metamagnetic transition.  As result,  the superconducting state recovers itself in much smaller field and can even merged 
 with  superconducting state in the small fields region.

The superconducting pairing is determined by the exchange interaction between the conduction electrons and the magnetic moments localised on uranium atoms.

In UCoGe the upward curvature of the upper critical field along c-axis is mostly determined by the longitudinal magnetic susceptibility decrease along with the magnetisation saturation.

In URhGe the superconducting state suppressed in field $H_y\approx 2 $T is recovered in fields interval $(9-13)$ T near the critical field. This phenomenon is related to the strong
increase of the pairing interaction caused mostly
by the strong augmentation of the magnetic susceptibility along $b$-direction in vicinity of the metamagnetic transition.

The nonmonotoneous behaviour of the upper critical field in UGe2 is explained by the strong increase of longitudinal magnetic susceptibility at the metamagnetic transition from FM1 to FM2.


\begin{figure}
\includegraphics
[height=.8\textheight]
{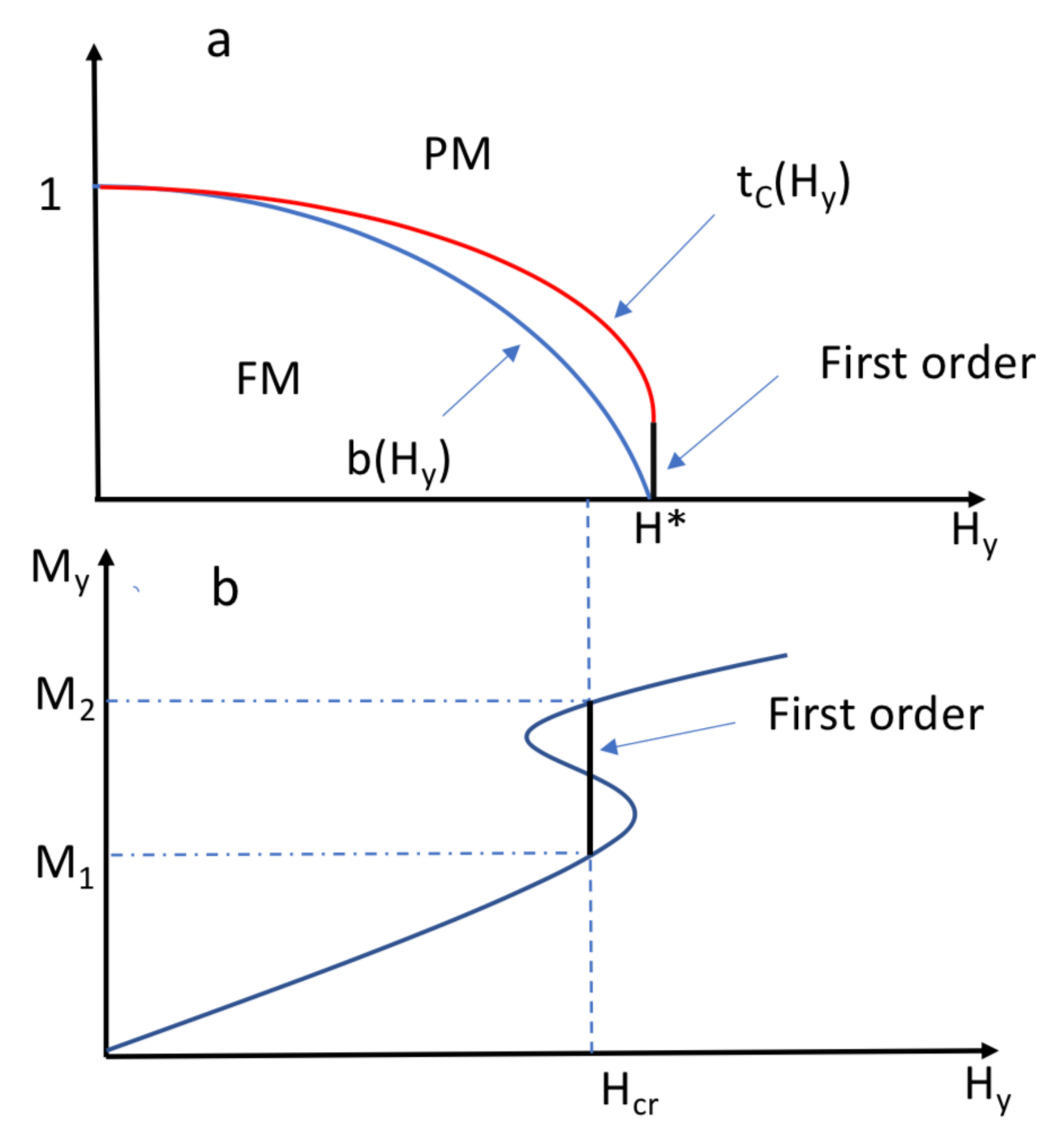}
 \caption{(Color online) 
a) Schematic behaviour of  the normalised Curie temperature $t_c(H_y)=\frac{T_c(H_y)}{T_{c0}}$ and coefficient $b(H_y)=\frac{\tilde\beta_{z}}{\beta_z}$. FM and PM stand for ferromagnetic and paramagnetic phases.
b) Schematic dependence $M_y(H_y)$ at $T<T_{cr}$ and $H_{cr}<H^\star$.}
\end{figure}

\begin{figure}
\includegraphics
[height=.8\textheight]
{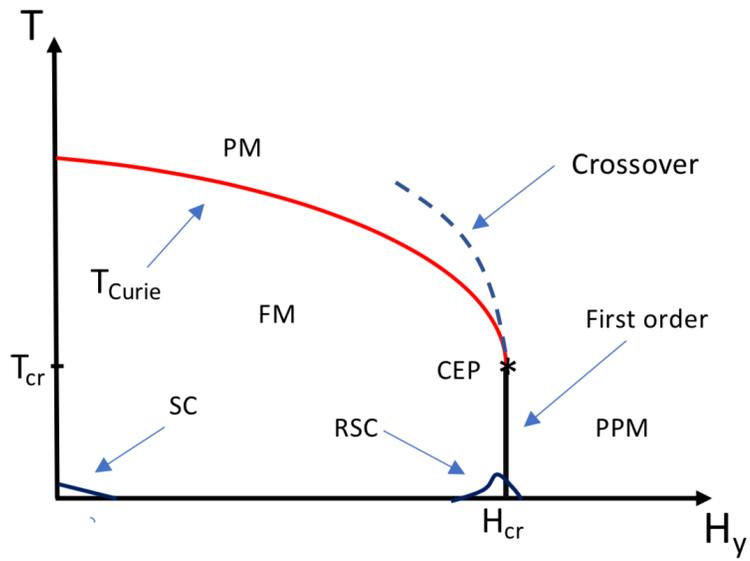}
 \caption{(Color online) 
Phase diagram UCoGe in magnetic field parallel to $b$-crystallographic direction. PM, FM, PPM denote paramagnetic, ferromagnetic and polarised paramagnetic phases. CEP is the critical end point.
SC and RSC are the superconducting and reentrant superconducting states.
}
\end{figure}

\begin{figure}
\includegraphics
[height=.8\textheight]
{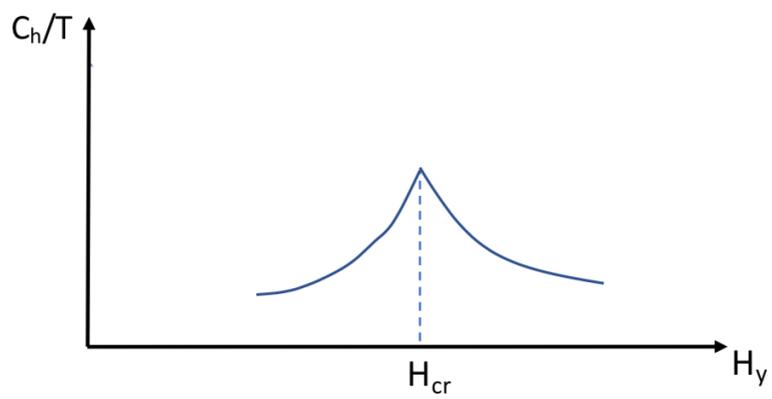}
 \caption{(Color online)
Schematic behaviour $C_h/T$ (See the main text).}
\end{figure}

\begin{figure}
\includegraphics
[height=.8\textheight]
{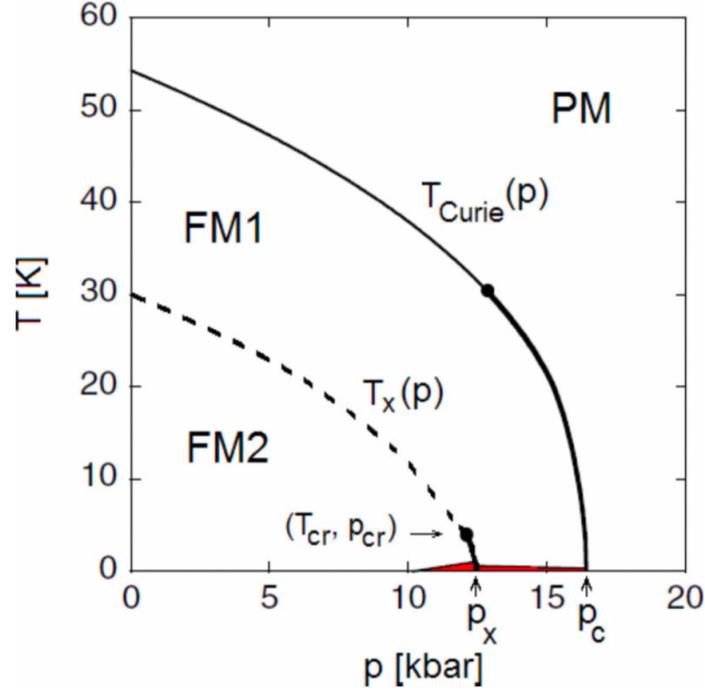}
\caption{(Color online) The schematic $P,T$ phase diagram of UGe$_2$ \cite{Huxley2002,Hardy2009}. Thick lines represent first-order transitions and thin lines denote second-order transitions. The dashed line indicates a crossover while the dots mark the positions of critical points. The superconducting region is represented in red area at bottom.
}
\end{figure}

\begin{figure}
\includegraphics
[height=.8\textheight]
{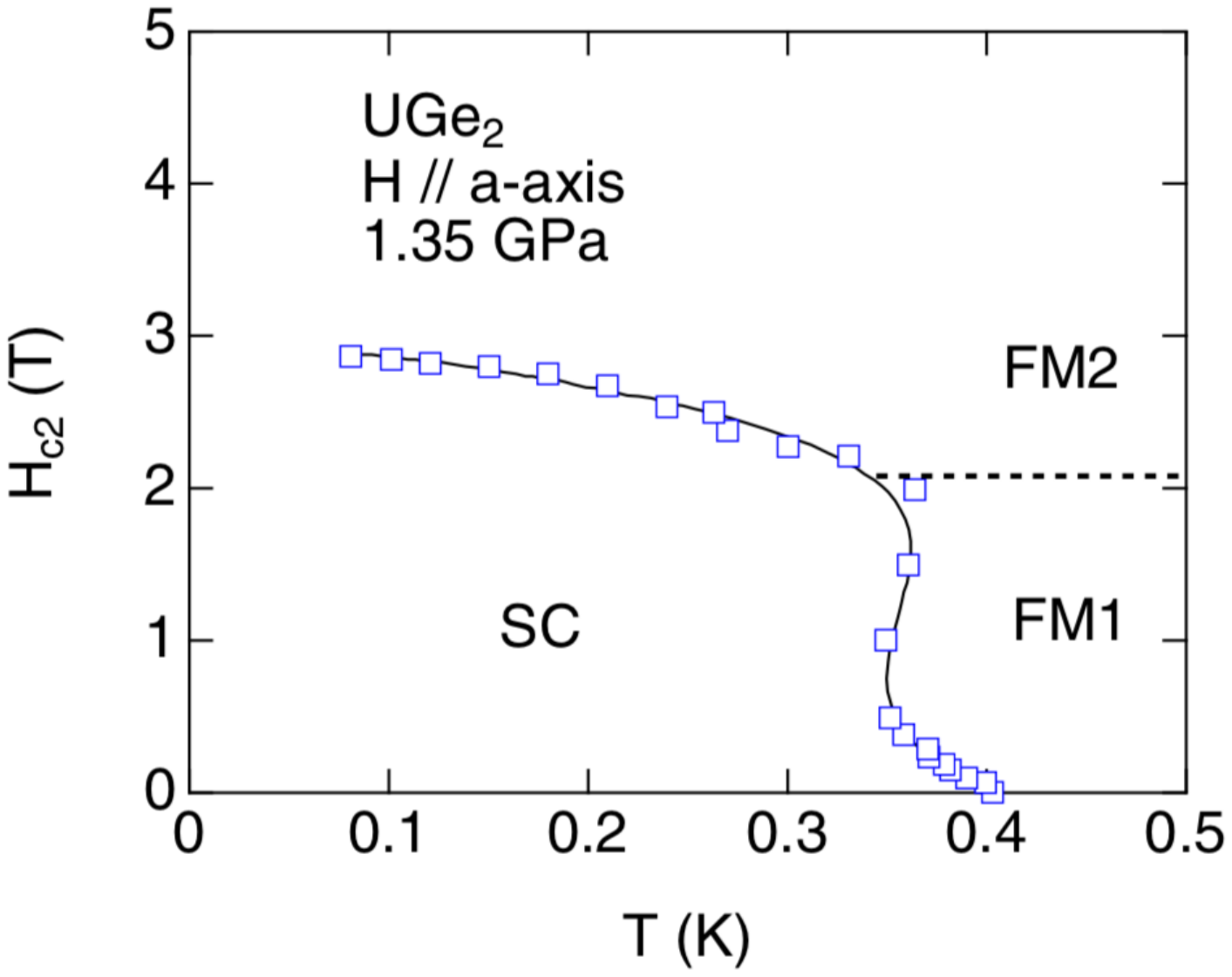}
\caption{(Color online) 
Temperature dependence of $H_{c2}$ for field parallel to $a$-axis in UGe$_2$ at 1.35GPa, which is just above $P_x$. The metamagnetic transition is detected at $H_x$ between FM1 and FM2 \cite{Huxley2001,Sheikin2001}.}
\end{figure}


\begin{thebibliography}{220}

\bibitem{Flouquet2019} D.Aoki, K.Ishida and J.Flouquet, J. Phys. Soc. Jpn. {\bf 88}, 022001 (2019).

\bibitem{Mineev2016}V.P.Mineev, Usp. Fiz. Nauk {\bf 187}, 129 (2017) [Phys.-Usp. {\bf 60}, 121 (2017).

\bibitem{Levy2005}  Levy F,  Sheikin I,  Grenier B, Huxley A D  {\it Science} {\bf 309} 1343 (2005)

\bibitem{Hardy2005}F. Hardy, A.D. Huxley, Phys. Rev. Lett. {\bf 94}, 247006 (2005).

\bibitem{Mineev2015} V.P.Mineev,  Phys. Rev. B {\bf 91} 014506 (2015).

\bibitem{Hardy2011} F.Hardy, D.Aoki, C.Meingast, P.Schweiss, P.Burger, H. v. L{\" o}hneysen, and J.Flouquet,
Phys.Rev.B {\bf 83}, 195107 (2011).

\bibitem{Nakamura2017}S.Nakamura, T.Sakakibara, Y.Shimizu, S.Kittaka, Y.Kono, Y.Haga, J.Pospisil, and E.Yamamoto, Phys. Recv. B {\bf 96},
094411 (2017).

\bibitem{YAoki1998}Y. Aoki, T.D. Matsuda, H. Sugawara, H. Sato, H. Ohkuni, R. Settai, Y. Onuki, E. Yamamot, Y. Haga, A.V. Andreev, V. Sechovsky, L. Havela, H. Ikeda,
K. Miyake, Journ.Mag.Magn.Mat. {\bf 177-181},271 (1998).


\bibitem{DAoki2011}D.Aoki, T.Combier, V.Taufour, T.D.Matsuda, G.Knebel, H.Kotegawa, and J.Flouquet,  J.Phys. Soc. Jpn. {\bf 80}, 094711 (2011).

\bibitem{Shick2002} A.B.Shick, Phys.Rev. B {\bf 65}, 180509(R) (2002).

\bibitem{Levitin1988} R.Z.Levitin, A.S.Markosyan, Usp. Fiz. Nauk {\bf 155}, 623 (1988) [Phys.-Usp. {\bf 31}, 730 (1988).

\bibitem{Sanchez2017} F.Wilhelm, J.P.Sanchez, J.-P.Brison, D.Aoki, A.B.Shick, and A.Rogalev, Phys.Rev. {\bf 95}, 235147 (2017).

\bibitem{F.Hardy2005}F. Hardy, A. Huxley, J. Flouquet, B. Salce, G. Knebel, D. Braithwaite,
D. Aoki, M. Uhlarz, C. Pfleiderer, Physica B {359-361}, 1111 (2005).

\bibitem{Miyake2009} A.Miyake, D.Aoki, and J. Flouquet, J.Phys. Soc. Jpn. {\bf 78}, 063703 (2009).


\bibitem{Braithwaite2016} D.Braithwaite, D.Aoki, J.-P.Brison, J.Flouquet, G.Knebel, A.Nakamura, and A.Pourret, Phys. Rev. Lett.
{\bf 120}, 037001 (2018).

\bibitem{Mineev2017} V.P.Mineev, Phys. Rev. B {\bf 95}, 104501 (2017).


\bibitem{StatPhys} L.D.Landau and E.M.Lifshitz, 
{\it Statistical Physics, Course of Theoretical Physics Vol V.} Oxford:  Butterworth-Heinemann,1995).

\bibitem{Miyake2008}  A.Miyake, D.Aoki, and J. Flouquet, J.Phys. Soc. Jpn. {\bf 77}, 094709 (2008).

\bibitem{Gourgout2016}A. Gourgout,  A. Pourret, G. Knebel,D. Aoki,G. Seyfarth,and J. Flouquet, Phys.Rev.Lett. {\bf 117}, 046401 (2016).

\bibitem{Keyes1958} R.W.Keyes, J. Phys. Chem. Solids {\bf 6}, 1 (1958).

\bibitem{Gantmakher} V.F.Gantmakher, I.B.Levinson, Zh. Eksp. Teor. Fiz. {\bf 74}, 261 (1978) [Sov. Phys. JETP {\bf 47} 133 (1978)].

\bibitem{Appel1978} J.Appel and A.W.Overhauser, Phys.Rev.B {\bf 18}, 758 {1978}.

\bibitem{Murzin}S.S.Murzin, S.I.Dorozhkin, A.C.Gossard, Pis'ma  Zh. Eksp. Teor.Fiz. {\bf 67}, 101 (1998)
[JETP Letters {\bf 67}, 113 (1998)].

\bibitem{Pal2012} H.K.Pal, V.I.Yudson, and D.L.Maslov, Lith.J.Phys. {\bf 52}, 142 (2012).


\bibitem{Prokes2002}K.Prokes, T.Tahara, Y.Echizen,T.Takabatake,T.Fujita, I.H.Hagmusa, J.C.P.Klaasse, E.Br{\"u}ck, F.R.deBoer,
M.Divis, V.Sechovsky, Physica B {\bf 311}, 220 (2002).


\bibitem{Kotegawa2015} H.Kotegawa, K.Fukumoto, T.Toyama, H.Tou, H.Harima, A.Harada, Y.Kitaoka, Y.Haga, E.Yamamoto, Y.Onuki, K.M.Itoh, and E.E.Haller, J.Phys. Soc. Jpn. {\bf 84}, 054710 (2015).

\bibitem{Tokunaga2015}Y. Tokunaga, D.Aoki, H.Mayaffre, S. Kr{\"a}mer, M.-H. Julien,C. Berthier, M. Horvati?,
H. Sakai, S. Kambe, and S. Araki, Phys. Rev. Lett. {\bf 114}, 216401 (2015).



\bibitem{McMillan1968} W.L.McMillan, Phys.Rev.{\bf 167}, 331 (1968).

\bibitem{Fay1980}D. Fay and J. Appel: Phys. Rev. B {\bf 22} (1980) 3173.

\bibitem{BE} W. F. Brinkman and S. Engelsberg, Phys. Rev. {\bf 169}, 417 (1968).

\bibitem{Chubukov2018}Yu.Sherkunov, A.V. Chubukov, and J.J.Betouras, Phys.Rev.Lett. {\bf 121}, 097001 (2018).

\bibitem{Yelland2011} E.A.Yelland, J.M.Barraclough, W.Wang, K.V. Kamenev, and A.D. Huxley, Nat. Phys. {\bf 7}, 890 (2011).


\bibitem{Rogalev2015}M.Taupin,  J.P.Sanchez, J.-P.Brison, D.Aoki, G.Lapertot,  F.Wilhelm, and A.Rogalev, Phys.Rev. {\bf 92}, 035124 (2015).

\bibitem{Hattori2013} R.Hattori, and H.Tsunetsugu, Phys. Rev. B {\bf 87}, 064501 (2013).

\bibitem{Karchev2003} N.Karchev, Phys.Rev.B {\bf 67}, 054416, (2003).


\bibitem{Mineev2020} V.P.Mineev, Annals of Physics (NY), to be published (2020).

\bibitem{Huy2008} N.Y.Huy, D.E.de Nijs, Y.K.Huang, and A.de Visser, Phys.Rev.Lett. {\bf 100}, 077002 (2008).

 \bibitem{Beilun2017} B.Wu, G.Bastien, M.Taupin, C.Paulsen, L.Howald, D.Aoki and J.-P. Brison, Nature Comm. {\bf 8}, 14480 (2017).

\bibitem{Ran-NatPhys} S.Ran, I-Lin Liu, YunSuk Eo, D.J.Campbell, P.M.Neves, W.T.Fuhrman, S.R.Saha, C.Eckberg, H.Kim, D.Graf, F.Balakirev, J.Singleton, J.Paglione and N.Butch, Nature Physics {\bf 15},1250 (2019).

\bibitem{Knebel2019}G.Knebel, W.Knafo, A.Pourret, Qun Niu, M.Valiska, D.Braithwaite, G.Lapertot, M.Nardone, A.Zitouni, S.Mishra, I.Sheikin, G.Seyfarth, J.-P.Brison, D.Aoki, J.Flouquet, J. Phys. Soc. Jpn {\bf 88}, 063707 (2019).

\bibitem{Pourret2020}Q.Niu, G.Knebel, D.Braithwaite, D.Aoki, G.Lapertot, M.Valiska, G.Seyfarth, W.Knafo, T.Helm, J.-P.Brison, J.Flouquet, and A.Pourret, 
arXiv:2003.08986 [cond-mat] (2020).


\bibitem{Knafo2012}W.Knafo, T.D.Matsuda, D.Aoki, F.Hardy, G.W.Scheerer, G.Ballon, M.Nardone, A.Zitouni, C.Meingast and J.Flouquet,
Phys. Rev. B {\bf 86}, 184416 (2012).

\bibitem{Troc2012} R.Troc, Z. Gajek, and A.Pikul, Phys. Rev. B {\bf 86}, 224403 (2012).

\bibitem{Mineev2013}V.P.Mineev, Phys. Rev. B {\bf 88}  224408 (2013).


\bibitem{Huxley2002} C.Pfleiderer and A.D.Huxley, Phys. Rev. Lett.
{\bf 89}, 147005 (2002).

\bibitem{Hardy2009}F. Hardy, C. Meingast, V. Taufour, J. Flouquet, H. v. L{\" o}hneysen, R. A. Fisher,
N. E. Phillips, A. Huxley, and J. C. Lashley, Phys.Rev.B {\bf 80} 174521 (2009).

\bibitem{Taufour2010} V. Taufour, D. Aoki, G. Knebel, and J. Flouquet, Phys. Rev. Lett. {\bf 105}, 217201 (2010).

\bibitem{Huxley2001}A.Huxley, I.Sheikin, E.Ressouche, N.Kernavanois, D.Braithwaite, R.Calemczuk, and J.Flouquet, Phys. Rev. B {\bf 63}, 144519 (2001).

\bibitem{Sheikin2001} I.Sheikin, A. Huxley, D. Braithwaite, J. P. Brison, S. Watanabe, K. Miyake, and J. Flouquet, Phys.Rev. B {\bf 64}, 220503(R) (2001).

\bibitem{Tateiwa2018} N.Tateiwa, Y.Haga, and E.Yamamoto, Phys.Rev.Lett. {\bf 121}, 237001 (2018).

\end{thebibliography}
\end{document}